\documentclass[a4paper,11pt,fleqn]{article}
\pdfoutput=1 
\usepackage[utf8]{inputenc}
\usepackage{jheppub} 
\usepackage{amsmath,amsfonts,amssymb}
\usepackage{graphicx}
\usepackage{mathtools}
\usepackage{mathrsfs}
\usepackage{bbm}
\usepackage{pifont}
\usepackage{marvosym}
\usepackage{units}
\usepackage{cancel}
\usepackage{hyperref}
\usepackage{enumerate}
\usepackage[small,loose,md,TABBOTCAP]{subfigure}
\usepackage[dvipsnames]{xcolor}
\usepackage{braket}
\usepackage[normalem]{ulem}
\usepackage{multirow} 
\usepackage{arydshln} 
\usepackage{euscript} 
\usepackage{wrapfig}
\usepackage{hyperref}

\newcommand{\Eqref}[1]{Equation~\eqref{#1}}
\newcommand{\Figref}[1]{Figure~\ref{#1}}
\newcommand{\Tabref}[1]{Table~\ref{#1}}
\newcommand{\Secref}[1]{Section~\ref{#1}}
\newcommand{\Appref}[1]{Appendix~\ref{#1}}

\newcommand{\eVdist}{\kern-0.06em}

\DeclareMathOperator{\re}{Re}
\DeclareMathOperator{\im}{Im}

\newcommand{\CenterObject}[1]{\ensuremath{\vcenter{\hbox{#1}}}}
\newcommand{\D}{\mathrm{d}}
\newcommand{\I}{\mathrm{i}}
\newcommand{\ChargeC}{\ensuremath{\mathcal{C}}}
\newcommand{\TimeT}{\ensuremath{\mathcal{T}}}
\newcommand{\ParityP}{\ensuremath{\mathcal{P}}}

\newcommand{\SO}[1]{\ensuremath{\mathrm{SO}(#1)}}
\newcommand{\SU}[1]{\ensuremath{\mathrm{SU}(#1)}}

\newcommand{\U}[1]{\ensuremath{\mathrm{U}(#1)}}
\newcommand{\Z}[1]{\ensuremath{\mathbbm{Z}_{#1}}} 

\newcommand{\T}[1]{\ensuremath{\mathrm{T}_{#1}}}
\newcommand{\SUCP}{\ensuremath{\mathrm{SU(3)}}--\,\ensuremath{\mathcal{CP}}}

\newcommand*{\rep}[2][]{\ensuremath{{\boldsymbol{#2}#1}}}
\renewcommand{\bar}[1]{\overline{#1}}

\newcommand{\CP}{\ensuremath{\ChargeC\ParityP}}

\newcommand{\elm}[1]{\mathsf{#1}}

\newcommand{\op}[1]{\ensuremath{\boldsymbol{\widehat{#1}}}}

\newcommand{\SusyNo}[0]{\texttt{SusyNo}}
\newcommand{\parityP}{\ensuremath{\EuScript P}}

\numberwithin{equation}{section}

\title{\boldmath \CP\ violation with an unbroken \CP\ transformation
\unboldmath}

\author[a,b]{Michael Ratz}
\author[c,1]{and Andreas Trautner\note{Corresponding author.}}

\affiliation[a]{Department of Physics and Astronomy, University of California,\\
Irvine, California 92697--4575, USA}
\affiliation[b]{On leave of absence from Physik Department T30, Technische
Universit\"at M\"unchen, \\ James--Franck--Stra\ss e~1, 85748 Garching, Germany}
\affiliation[c]{Bethe Center for Theoretical Physics und Physikalisches Institut der Universit\"at Bonn,\\
Nussallee 12, 53115 Bonn, Germany}

\emailAdd{mratz@uci.edu}
\emailAdd{atrautner@uni-bonn.de}

\preprint{UCI-2016-26}

\abstract{A \CP\ conserving \SU3 gauge theory is spontaneously broken to \T7 by
the vacuum expectation value (VEV) of a \rep{15}--plet.  Even though the
\SUCP\ transformation is not broken by the VEV,  the theory exhibits
physical \CP\ violation in the broken phase. 
This is because the \SUCP\ transformation corresponds to the unique order--two outer
automorphism of \T7, which is not a physical \CP\ transformation for the \T7
states, and there is no other possible \CP\ transformation.
We explicitly demonstrate that \CP\ is violated by calculating a \CP\
odd decay asymmetry in the broken phase. This scenario provides us
with a natural protection for topological vacuum terms, ensuring that
$\theta\,G_{\mu\nu}\widetilde{G}^{\mu\nu}$  is absent even though \CP\ is
violated for the physical states of the model. \\
 \vspace{3cm} }

\notoc
\begin{document}
\maketitle

\section{Introduction}
\label{sec:intro0}

Violation of the combined transformation of charge conjugation and parity
(\CP{}) has been observed in decays and oscillations of K and B mesons
\cite{Christenson:1964fg, Agashe:2014kda}. Complementary to that, violation of
the time reversal transformation ($\TimeT$) is implied by the \CP\TimeT\ theorem
and has been verified independently \cite{Lees:2012}. Likewise, recent global
fits of neutrino oscillation data point towards \CP\ violation also in the
lepton sector \cite{Esteban:2016qun}. Given that the measurement of leptonic
mixing angles has entered its precision phase, and anticipating the near future
measurement of neutrino masses, the experimental efforts to pin down all
parameters which constitute the SM flavor puzzle is close to being completed.

\medskip

Even though the observed fermion masses, mixings, and \CP\ violation can be
consistently parametrized in the framework of \mbox{$3\times3$} (CKM
\cite{Kobayashi:1973fv} or PMNS \cite{Maki:1962mu}) unitary mixing, the overall
theoretical situation is in many ways unsatisfactory. First of all, $\CP$ violation
beyond the standard model is necessary in order to explain the baryon asymmetry
of the universe \cite{Sakharov:1967dj}. Furthermore, \CP\ violation is observed
only in flavor changing transitions mediated by the weak interaction, not in
strong interactions, thereby giving rise to a severe fine--tuning problem of the
$\overline{\theta}_\mathrm{QCD}$ parameter. Ultimately, the sheer multitude of
parameters in the flavor sector and their distinct pattern cries out for an
underlying fundamental explanation. Therefore, more theoretical work is
necessary in order to unveil the underlying structure of the flavor puzzle,
including the origin of \CP\ violation.

\medskip

From a formal point of view, \CP\ transformations are quite particular. While
\ParityP\ or \ChargeC\ transformations individually relate otherwise
disconnected representations of the space--time and gauge symmetries of a model,
a \CP\ transformation maps each irreducible representation (irrep) to \emph{its
own} complex conjugate representation. This implies that in theories with a real
valued action, \CP\ cannot be broken \emph{maximally} by simply leaving out or
adding fields, in vast contrast to, for example, the parity transformation in a
chiral theory. From a group--theoretical point of view this is reflected by the
fact that \CP\ transformations are special outer\footnote{The automorphism must
be outer if the corresponding symmetry group has complex representations.}
automorphisms of the continuous \cite{Grimus:1995zi}, discrete
\cite{Holthausen:2012dk,Chen:2014tpa} and space--time \cite{Buchbinder:2000cq}
symmetries of a model, which map the irreps of each group to their own complex
conjugate representations \cite{Trautner:2016ezn}. This should be contrasted to
other outer automorphisms, such as parity or charge conjugation for example,
which map irreps to other irreps that may not be present in a model to begin
with.

\medskip

This notion defines \CP\ transformations as special automorphisms of symmetry
groups. However, it is not guaranteed that such an automorphism exists for a
given symmetry group \cite{Chen:2014tpa}. While it is known that \CP\ outer
automorphisms exist for simple Lie groups \cite{Grimus:1995zi} and the
Poincar\'e group \cite{Buchbinder:2000cq} it has been pointed out that there are
certain discrete groups which violate \CP\ by the intrinsic complexity of their
Clebsch--Gordan coefficients \cite{Chen:2009gf}. In more detail, these
so--called ``type I'' groups prohibit simultaneous
$\rep[_i]{r}\leftrightarrow\rep[_i]{r}^*$ (for all $i$, labeling the irreps of
the corresponding group) transformations \cite{Chen:2014tpa}. That is, these
type I groups do not allow for outer automorphisms which could be identified
with physical \CP\ transformations for all of their irreps. If a model features
such a group as global symmetry and a sufficiently large number of irreps,
then \CP\ transformations are not possible in consistency with the symmetry
group of the model. Explicitly it has been found that the necessarily complex
Clebsch--Gordan coefficients of type I groups then enter \CP\ odd basis
invariants, thereby giving rise to particle--antiparticle asymmetries in
oscillations and decays \cite{Chen:2014tpa}. Due to the fact that the arising
physical \CP{}--odd complex phases are discrete and calculable, this phenomenon
has been termed explicit geometrical \CP\ violation \cite{Chen:2009gf,
Branco:2015hea}.

\medskip

In this context, an important fact is that type I groups can arise as subgroups
of simple Lie groups. This gives rise to the puzzling situation in which a Lie
group $G$ allows for a perfectly well--defined \CP\ transformation, i.e.\ a
particular complex conjugation outer automorphism, whereas the type I subgroup,
$H\subset G$, does not. Thus, \CP\ is not conserved at the level of $H$, and the
question arises when and how \CP\ is broken in a possible breaking of
$G\rightarrow H$. Na\"{i}vely, one might expect that in a dynamical setting it
should be the vacuum expectation value (VEV) which gives rise to \CP\ violation.
Rather surprisingly, it turns out  that this is not necessarily the case. We
will show that the VEV which spontaneously breaks $G\rightarrow H$ does
\emph{not} break the complex conjugation outer automorphism, i.e.\ the VEV is
\CP\ conserving. Nevertheless, physical \CP\ is violated at the level of $H$,
and we will substantiate this claim by an explicit calculation of non--vanishing
\CP\ odd basis invariants that give rise to a physical decay asymmetry. The way
the conundrum is  resolved is the following: The conserved outer automorphism
which gives rise to \CP\ conservation at the level of $G$ is conserved by the
VEV and, hence, also a conserved outer automorphism at the level of $H$.
Nevertheless, at the level of $H$ this outer automorphism is no  complex
conjugation automorphism. Thus, once the physical states of a theory are $H$
states, the conserved outer automorphism can no longer be interpreted as a
physical \CP\ transformation. An anticipated distinct outer automorphism
transformation, which \emph{would} correspond to a physical \CP\ transformation
at the level of $H$, is prohibited by the group structure of $H$, and it would
also not be a consistent automorphism of $G$ to begin with.

\medskip

The investigation in this paper is based on an economic toy example. We use an
\SU3 gauge theory which is broken to the type I group \T7 by a the VEV of a
complex scalar $\phi$, transforming in the \rep{15}--plet representation of \SU3
with the Dynkin indices $(2,1)$. We assume physical \CP\ conservation at the
level of \SU3 and show that the VEV does not break the corresponding outer
automorphism. Yet we will find \CP\ violating decays of a physical scalar to
gauge bosons in the broken phase.

\medskip

The rest of the paper is organized as follows. In \Secref{sec:Model}, we
define our model. \Secref{sec:SU3toT7} details the spontaneous breaking of \SU3
to \T7. In \Secref{sec:PhysicalStates}, we discuss how the \T7 states are
related to the states of the \SU3 theory. In \Secref{sec:CPviolation}, we show
explicitly that \CP\ is broken in the \T7 phase. \Secref{sec:ThetaProtection}
contains a comment on the $\theta$ term of the \SU3 theory. Finally,
\Secref{sec:Summary} contains our conclusions. Some details are deferred to the
appendices.

\section{The model}
\label{sec:Model}

We consider an \SU3 gauge theory with the Lagrangean
\begin{equation}\label{eq:Lagreangean}
\mathscr{L}~=~\left(D_\mu\,\phi\right)^\dagger\left(D^\mu\,\phi\right)
-\frac{1}{4}\,G^a_{\mu\nu}\,G^{\mu\nu,a}-\mathscr{V}(\phi)\;,
\end{equation}
with $D_\mu=\partial_\mu-\I g\,A_\mu$ being the ordinary gauge covariant
derivative and $G_{\mu\nu}^a$ the field strength tensors. The complex
scalar $\phi$ is charged under \SU3, transforming in the \rep{15}--plet
representation. The scalar potential is given by
\begin{equation}\label{eq:potential}
 \mathscr{V}(\phi)~=~-\mu^2\,\phi^\dagger\phi+\sum^5_{i=1}\lambda_i\,\mathcal{I}^{(4)}_i(\phi)\;,
\end{equation}
where we take $\mu^2>0$, and the $\mathcal{I}^{(4)}_i$ denote the five
independent completely symmetric quartic \SU3 invariants in the contraction
$\rep{\bar{15}}\otimes\rep{15}\otimes\rep{\bar{15}}\otimes\rep{15}$. All
other invariant vanish. These invariants are real, which allows us to take $\lambda_i\in\mathbbm{R}$ without loss of generality.
Further details on the derivation of the invariants can be found in \Appref{app:Invarians}.
There is a region in the parameter space in which this
potential has a global minimum which gives rise to a VEV $\langle\phi\rangle$
that spontaneously breaks $\SU3\rightarrow\T7$ \cite{Luhn:2011ip, Merle:2011vy}.
The linear symmetry of the vacuum is \T7, a discrete group with $21$ elements.
The physical spectrum after spontaneous symmetry breaking (SSB) is given in
\Tabref{tab:physicalFields}. We will give further details of \T7, including its
embedding in \SU3 and branching rules in \Secref{sec:Embedding}.

\begin{table}[t]
 \centering
 \begin{tabular}{cc:cc}
 Name & \multicolumn{2}{l}{$\SU3\xrightarrow{\langle\phi\rangle}~$ Name} & $\T7$  \\
 \hline
  \multirow{2}{*}{$A_\mu$} & \multirow{2}{*}{\rep{8}} & $Z_\mu$ &
  \;\rep{1_1}  \\
  &                                                 & $W_\mu$ & \rep{3}   \\
  \hline
  \multirow{5}{*}{$\phi$} & 
  \multirow{5}{*}{\hspace{11pt}$\rep{15}$\hspace{11pt}} &
  $\re\sigma_0,\im\sigma_0$ & \;\rep{1_0}  \\
  &                                                                             
       & $\sigma_1$ & \;\rep{1_1}  \\
  &                                                 				      & $\tau_1$   & \rep{3}      \\
  &                                                 				      & $\tau_2$   & \rep{3}      \\
  &                                                 				      & $\tau_3$   & \rep{3}     \\
  \hline
\end{tabular}
\caption{Physical fields before and after the symmetry breaking.
The number of real degrees of freedom before and after SSB coincide as $16+30=24+22$.}\label{tab:physicalFields}
\end{table}

\medskip

The action derived from \eqref{eq:Lagreangean} is automatically invariant under the simultaneous outer automorphisms of \SU3 and the Lorentz group under which
the gauge and scalar fields transform as
\begin{subequations}
\begin{align}\label{eq:CPTrafo}
A^a_\mu(x)~&\mapsto~R^{ab}\,\parityP_\mu^{\;\nu}\,A^b_\nu(\parityP\,x)\;, \\
\phi_i(x)~&\mapsto~U_{ij}\,{\phi_j^*}(\parityP\,x)\;.
\end{align}
\end{subequations}
Here $\parityP=\mathrm{diag}(1,-1,-1,-1)$ is the usual spatial reflection, while $R$ and $U$ are
the representation matrices of the outer automorphism of \SU3 fulfilling
\begin{equation}\label{eq:ConsistencyCondition}
R_{aa'}\,R_{bb'}\,R_{cc'}\,f_{a'b'c'}~=~f_{abc}\;,
\quad\text{and}\quad 
U\,\left(-\mathsf{t}^\mathrm{T}_a\right)\,U^{-1}~=~R_{ab}\,\mathsf{t}_b\;.
\end{equation}
Here, $\mathsf{t}_a$ and $f_{abc}$, with the usual relation $\left[\mathsf{t}_a,\mathsf{t}_b\right]=\I\, f_{abc}\,\mathsf{t}_c$, are the
generators and structure constant of the Lie algebra, respectively. At the level
of \SU3, this outer automorphism maps all irreps to their own complex conjugate
and, therefore, is the most general possible physical \CP\ transformation. The
transformation is conserved by the action implying that the model is \CP\
symmetric in the unbroken phase. In addition, the VEV fulfills
$U\,\langle\phi\rangle^*=\langle\phi\rangle$ and, therefore, does not break the
outer automorphism. Surprisingly, we will find that even though the
\SUCP\ transformation is not broken, physically, \CP\ is no
longer conserved once \SU3 gets broken to \T7.

\medskip

As will be detailed below, the unique \Z2 outer automorphism of \SU3
corresponds to the unique \Z2 outer automorphism of \T7. Therefore, the actual symmetry breaking chain of the model is given by
\begin{equation}
  \SU3\rtimes\Z2~\xrightarrow{\langle\phi\rangle}~\T7\rtimes\Z2\;.
\end{equation}
Crucially, the \Z2 transformation on the left--hand side corresponds to a \CP\ transformation, 
while the identical transformation on the right--hand side cannot be interpreted as \CP. 
In more detail, at the level of \T7, the unique \Z2 outer automorphism acts as (cf.\ also
\Tabref{tab:CharacterTable})%
\begin{equation}\label{eq:T7Out}
 \mathrm{Out}(\T7)~:\qquad
\rep{1_1}~\longleftrightarrow~\rep{1_1}\;,\quad
\rep{\bar{1}_1}~\longleftrightarrow~\rep{\bar{1}_1}\;,\quad
\rep{3}~\longleftrightarrow~\rep{\bar{3}}\;.
\end{equation}%
\begin{table}[t!]
 \centering
 \includegraphics{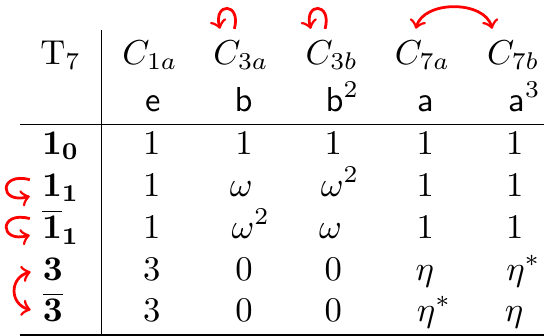}
\caption{Character table of \T7.
The arrows illustrate the action of the unique \Z2 outer automorphism of \T7.
We use the definitions $\omega:=\mathrm{e}^{2\pi \I/3}$ and $\eta=\rho+\rho^2+\rho^4$ with $\rho:=\mathrm{e}^{2\pi \I/7}$.}
\label{tab:CharacterTable}
\end{table}%
This transformation does \emph{not} map all \T7 irreps to their own complex
conjugate representation. Therefore, it does not correspond to a physical \CP\
transformation if triplet and non--trivial singlet representations of \T7 are
present simultaneously.
This is the case in the given model. That is, the model does not allow
for a physical \CP\ transformation at the level of \T7.
Hence, the setting exhibits \CP\ violating processes once
\SU3 gets broken to \T7. For definiteness, we will explicitly show in
\Secref{sec:CPviolation} that there is a \CP\ asymmetry in the decay of a
heavy charged scalar to massive gauge bosons of the broken \SU3 symmetry.

\medskip

Here it should also be noted that any supposed physical \CP\ transformation
$\rep{r}^{\phantom{*}}_{\mathrm{T}_7}\leftrightarrow\rep{r}_{\mathrm{T}_7}^*$ is
inconsistent with the structure of \T7. For this reason the group has been
classified as a finite group of ``type I'' in \cite{Chen:2014tpa}. In fact,  a
supposed transformation
$\rep{r}^{\phantom{*}}_{\mathrm{T}_7}\leftrightarrow\rep{r}_{\mathrm{T}_7}^*$ is
not a consistent automorphism at the level of \SU3 to begin with. Imposing this
transformation as a symmetry nonetheless, enforces $g=\lambda_i=0$,  i.e.\
forbids all interactions.

\section{\boldmath\texorpdfstring{$\SU3$}{} and \texorpdfstring{$\T7$}{}
subgroup\unboldmath}
\label{sec:SU3toT7}

\subsection{Embedding}
\label{sec:Embedding}

The discrete group \T7 can be generated by two elements with the presentation
\begin{equation}
\Braket{\elm{a}, \elm{b}\,|\, \elm{a}^7~=~\elm{b}^3~=~\elm{e}\;, 
\elm{b}^{-1}\,\elm{a}\,\elm{b}~=~\elm{a}^4 }\;.
\end{equation}
For the triplet representation we choose a basis in which $\elm{a}$ and
$\elm{b}$ are represented by
\begin{equation}\label{eq:T7gens}
 A~=~\begin{pmatrix} \rho & 0 & 0 \\ 0 & \rho^2 & 0 \\ 0 & 0 & \rho^4 \end{pmatrix}
 \quad \text{and}\quad
 B~=~\begin{pmatrix} 0 & 1 & 0 \\ 0 & 0 & 1 \\ 1 & 0 & 0 \end{pmatrix}\;,
\end{equation}
respectively, with $\rho:=\mathrm{e}^{2\pi \I/7}$.

\medskip

The group elements of \SU3 for an arbitrary representation \rep{r} are given by
\begin{equation}\label{eq:SU3GroupElement}
X^{(\rep{r})}~=~\exp\left(\I\,\alpha_a\,\mathsf{t}^{(\rep{r})}_a\right)\;,
\end{equation}
where $\mathsf{t}^{(\rep{r})}_a$ denote the generators of the representation
\rep{r} and $\alpha_a$ the parameters. We emphasize that there are two generally
different spaces which are relevant here. While both, $[X^{(\rep{r})}]_{ij}$ and
$[\mathsf{t}^{(\rep{r})}_a]_{ij}$, live in the ``$ij$--space'' there is also the
``$a$'' or adjoint space in which the parameters $\vec\alpha$ as well as the
gauge bosons live. This is important, because it is possible to choose bases
independently for each of these spaces.\footnote{Only for special
transformations --- which are precisely the group transformations --- it is
possible to compensate transformations of the $ij$--space by transformations of
the adjoint space.}

\medskip

For practical reasons, we do not work in the standard Gell--Mann basis of the adjoint space but follow the
basis choice of Fonseca's \texttt{Mathematica} package \SusyNo~\cite{Fonseca:2011sy}.
The generators of the fundamental representation $\rep{r}=\rep{3}$ used in this work are specified in \Appref{App:SU3}.

\medskip

The discrete subgroup \T7 is embedded into \SU3 via the irreducible triplet
representation. In the given basis we find that $A$ and $B$ are obtained from
\eqref{eq:SU3GroupElement} by the choice of parameters
\begin{equation}\label{eq:T7ParametersAB}
 \vec\alpha^{(A)}~=~\frac{2 \pi}{7}\left(0,0,0,0,0,0,\sqrt{3},5\right)
 \quad\text{and}\quad
 \vec\alpha^{(B)}~=~\frac{4 \pi}{3\sqrt3}\left(0,0,1,1,1,0,0,0\right)\;.
\end{equation}
The branching rules of representations under $\SU3\to\T7$ can be calculated with the help of \cite{Fallbacher:2015pga}.
Branchings relevant to this work are
\begin{subequations}
\begin{align}\label{eq:BranchingRules8}
\rep{8}~&\to~\rep{1_1}\,\oplus\,\rep{\bar{1}_1}\,\oplus\,\rep{3}\,\oplus\,\rep{\bar{3}}\;,\\ \label{eq:BranchingRules15}
 \rep{15}~&\to~\rep{1_0}\,\oplus\,\rep{1_1}\,\oplus\,\rep{\bar{1}_1}\,\oplus\,\rep{3}\,\oplus\,\rep{3}\,\oplus\,\rep{\bar{3}}\,\oplus\,\rep{\bar{3}}\;.
\end{align}
\end{subequations}

\subsection{Outer automorphism}
\label{sec:Out}

Fonseca's basis choice has the virtue of being a \CP\ basis \cite{Grimus:1995zi}, meaning that
$U=\mathbbm{1}$ in \eqref{eq:ConsistencyCondition} irrespective of the specific representation.
At the same time, the outer automorphism transformation in the adjoint space in this basis is given by
\begin{equation}\label{eq:Rmatrix}
 R~=~\mathrm{diag}(-1,-1,1,1,1,-1,-1,-1)\;,
\end{equation}
which can easily be computed from \eqref{eq:ConsistencyCondition}.
We stress that in contrast to the $U$'s, it is not possible to chose a basis in
which $R$ is trivial, for a non--trivial automorphism.

\medskip

Since we ultimately will break \SU3 to \T7, it is convenient to rotate the
$ij$--space of the scalar representation to a \T7--diagonal basis, in which
\eqref{eq:BranchingRules15} is explicitly realized for the \T7 generators
parametrized by \eqref{eq:T7ParametersAB}. The
corresponding matrix for the basis change is given in \Appref{App:SU3}. Most
importantly, the transformation matrix of the outer automorphism, $U$ (cf.\
\Eqref{eq:ConsistencyCondition}), is not invariant under such a basis
change.\footnote{Note that $U$ under basis changes rotates with
$V\,U\,V^\mathrm{T}$ rather than $V\,U\,V^\dagger$.} In particular, we find
that
\begin{align}\label{eq:UCP}
 U^{(\T7)}_{\rep{15}}~&=~\mathbbm{1}_1\,\oplus\,\begin{pmatrix}0 & 1 \\ 1 & 0 \end{pmatrix}\,\oplus\,\mathbbm{1}_{12}\;,
\end{align}
where $\mathbbm{1}_n$ denotes the $n$--dimensional unit matrix. This shows that the \SU3 outer automorphism acts like
\eqref{eq:T7Out} on the \T7 representations.

\medskip

Analogously, it is possible to choose a basis for the gauge bosons in the $a$--space such that \eqref{eq:BranchingRules8}
is explicitly realized. The corresponding basis change is also given in \Appref{App:SU3}.
We will see later that this basis also corresponds to the physical basis for the gauge bosons.

\medskip

Using the basis changes in reverse, one can also show that the na\"{i}ve
\CP\ transformation at the level of \T7 (which would have to take
$\rep[_i]{r}\rightarrow\rep[_i]{r}^*$ $\forall$ \T7 irreps $i$) does not fulfill
\eqref{eq:ConsistencyCondition} and, hence, does not correspond to a valid \SU3
automorphism.

\section{Physical states in the \texorpdfstring{$\boldsymbol{\T7}$}{} phase}
\label{sec:PhysicalStates}

In the \T7--diagonal basis of \SU3, and using unitary gauge, the scalar $\phi$ can be written as
\begin{equation}\label{eq:unitaryGauge}
 \phi~=~\left(v\,+\,\phi_1,\,\frac{\phi_2}{\sqrt{2}},\,
 \frac{\phi_2^*}{\sqrt{2}},\,\phi_4,\,\phi_5,\,\phi_6,\,\frac{\phi_7}{\sqrt{2}},
 \,\frac{\phi_8}{\sqrt{2}},\,
 \frac{\phi_9}{\sqrt{2}},\,\phi_{10},\,\phi_{11},\,\phi_{12},
 \,\frac{\phi_7^*}{\sqrt{2}},\,\frac{\phi_8^*}{\sqrt{2}},\,
 \frac{\phi_9^*}{\sqrt{2}}  \right)\;,
\end{equation}
featuring $22=30-8$ real degrees of freedom.
In this basis, it is straightforward to identify the \T7 representations of
the components as
\begin{align}\label{eq:T7fields}\nonumber
 \phi_1~&\mathrel{\widehat{=}}~\rep[_0]{1}\;,& \phi_2~&\mathrel{\widehat{=}}~\rep[_1]{1}\;,& \\ \nonumber
 T_1~&:=~\left(\phi_4,\,\phi_5,\,\phi_6 \right)~\mathrel{\widehat{=}}~\rep{3}\;,& T_2~&:=~\left(\phi_7,\,\phi_8,\,\phi_9\right)~\mathrel{\widehat{=}}~\rep{3}\;,& \\
 \bar{T}_3~&:=~\left(\phi_{10},\,\phi_{11},\,\phi_{12}\right)~\mathrel{\widehat{=}}~\rep{\bar{3}}\;.&
\end{align}
The VEV in this basis is simply given by
\begin{equation}
 \langle\phi\rangle_1~=~v \quad\text{and}\quad \langle\phi\rangle_i~=~0\quad\text{for}\quad i=2,\dots,15\;.
\end{equation}
Minimizing the potential one finds that
\begin{equation}\label{eq:vvsmu}
 |v|~=~\mu\times3\,\sqrt{\frac72}\left(-7\sqrt{15}\,\lambda_1+14\sqrt{15}\,\lambda_2+20\sqrt{6}\,\lambda_4+13\sqrt{15}\,\lambda_5\right)^{-1/2}\;.
\end{equation}
In what follows, we will choose $v$ to be real and positive without loss of
generality. The physical \T7 states of the gauge bosons are complex linear
combinations of the $A^\mu_a$'s. The massive and \T7 charged gauge bosons are
given by
\begin{subequations}\label{eq:GaugeBosons}
\begin{align}
 Z^{\mu}~&=~\frac{1}{\sqrt{2}}\left(A^{\mu}_7-\I\,A^{\mu}_8\right)\;,\label{eq:Z}& \\
 W^{\mu}_1~&=~\frac{1}{\sqrt{2}}\left(A^{\mu}_4-\I\,A^{\mu}_1\right)\;,&\\
 W^{\mu}_2~&=~\frac{1}{\sqrt{2}}\left(A^{\mu}_5-\I\,A^{\mu}_2\right)\;,&\\
 W^{\mu}_3~&=~\frac{\I}{\sqrt{2}}\left(A^{\mu}_6-\I\,A^{\mu}_3\right)\;,&
\end{align}
\end{subequations}
They obtain masses
\begin{equation}
 m_Z^2~=~\frac73\,g^2\,v^2\quad\text{and}\quad m_W^2~=~g^2\,v^2\;.
\end{equation}
The physical scalars arising from $\phi$ are mixtures of the fields listed in \eqref{eq:T7fields}.
For the one--dimensional representations one finds
\begin{subequations}
\begin{align}
 \re\sigma_0~&=~\frac{1}{\sqrt{2}}\left(\phi_1+\phi_1^*\right)\;,& \im\sigma_0~&=~-\frac{\I}{\sqrt{2}}\left(\phi_1-\phi_1^*\right)\;,& \\
 \sigma_1~&=~\phi_2\;,&
\end{align}
\end{subequations}
with masses
\begin{subequations}
\begin{align}
 m_{\re\sigma_0}^2&~=~2\,\mu^2\;, \qquad m_{\im\sigma_0}^2~=~0\;, \\ \label{eq:MSigma1}
 m_{\sigma_1}^2&~=~-\,\mu^2+\,\sqrt{15}\,\lambda_5\,v^2\;.
\end{align}
\end{subequations}
The massless mode can be understood noting that $\im\sigma_0$ is the Goldstone
boson of an additional global \U1 symmetry of the potential \eqref{eq:potential}
which is spontaneously broken by $\langle\phi\rangle$. This symmetry prohibits a
possible cubic coupling term for $\phi$. The massless mode can be avoided by
softly breaking the \U1 via a reintroduction of the cubic term.  Alternatively
one could also gauge the additional \U1 upon which the would--be Goldstone boson
$\im\sigma_0$ gets eaten by the \U1 gauge boson. Either way, this mode does not
play any role in our discussion.

\medskip

In contrast to the one dimensional representations, the triplet representations
appear in identical copies. Therefore, the physical states are mixtures of
$T_1$, $T_2$, and $\bar{T}_3^*$, with mixing parameters depending on the
potential parameters $\lambda_i$. The physical states are given by
\begin{equation}
 \begin{pmatrix}
  \tau_1 \\ \tau_2 \\ \tau_3
 \end{pmatrix}
 ~=~
 \underbrace{
 \begin{pmatrix}
  V_{11} & V_{12} & V_{13}\\
  V_{21} & V_{22} & V_{23}\\
  V_{31} & V_{32} & V_{33}\\
 \end{pmatrix}}_{=~V}\,
 \begin{pmatrix}
  T_2 \\ \bar{T}_3^*\\ T_1
 \end{pmatrix}
 \;.
\end{equation}
The mixing matrix $V$ is an orthogonal matrix with entries depending on the
potential parameters $V_{ij}=V_{ij}(\lambda_{k})$ which are too complicated to
be displayed here.\footnote{While the mixing of physical states is dependent on
the potential parameters, we note that the composition of the unphysical
Goldstone bosons is independent of the potential parameters. This allows for the
general choice \eqref{eq:unitaryGauge}.} Instead of discussing the most general
case, we will settle to a specific set of potential parameters, which is
sufficient to prove our point. We choose the parameter values\footnote{One
should not be bothered by the fact that $\lambda_{3,4}=0$, as there are also
global \T7 minima for parameter choices $\lambda_i\neq0~\forall i$.}
\begin{align}\label{eq:parameters}
 \lambda_1~&=~0.1 \;,& \lambda_2~&=~-0.2 \;,& \lambda_3~&=~0 \;,& \lambda_4~&=~0 \;,& \lambda_5~&=~1 \;,&
\end{align}
which lead to a global \T7 minimum of the potential \eqref{eq:potential} with
a VEV $v^2\approx\mu^2\times0.856$. The corresponding mixing matrix of the
physical states is given by
\begin{equation}
V~\approx~
\begin{pmatrix}
  \phantom{-}0.950 & \phantom{-}0.288 & 0.121 \\
 -0.304 & \phantom{-}0.941 & 0.146 \\
 -0.072 & -0.175 & 0.982 \\
 \end{pmatrix}\;,
\end{equation}
and the masses of the physical states are
\begin{align}
m^2_{\tau_1}~&\approx~\mu^2\,\times\,0.946 \;,&
m^2_{\tau_2}~&\approx~\mu^2\,\times\,0.322 \;,&
m^2_{\tau_3}~&\approx~\mu^2\,\times\,0.142 \;,&
\end{align}
as well as
\begin{equation}
m_{\sigma_1}^2~\approx~\mu^2\,\times\,2.316\;.
\end{equation}%
The hierarchies appearing here originate from the mild hierarchies in
$\lambda_{1,2}/\lambda_5$.

\section{Physical \boldmath\CP\ violation in the \texorpdfstring{$\T7$}{} phase\unboldmath}
\label{sec:CPviolation}

Given the physical fields, the stage is set for an explicit proof of physical
\CP\ violation in the broken phase. The conserved outer automorphism, which
corresponds to the \CP\ transformation at the level of \SU3, acts on the
physical states as
\begin{equation}\label{eq:SU3CP}
\mathrm{Out}(\T7)~:\;
\begin{array}{lll}
    Z_\mu(x)~\mapsto~-\parityP_\mu^{\;\nu}\,Z_\nu(\parityP\,x)\;, & \sigma_0(x)~\mapsto~\sigma_0(\parityP\,x)\;, &  \\
    W_\mu(x)~\hspace{-2.75pt}\mapsto~\hspace{13pt}\parityP_\mu^{\;\nu}\,W^*_\nu(\parityP\,x)\;, & \sigma_1(x)~\mapsto~\sigma_1(\parityP\,x)\;, & \tau_i(x)~\mapsto~\tau_i^*(\parityP\,x)\;.
\end{array}
\end{equation}
Clearly, this does not correspond to a physical \CP\ transformation. 

\medskip

This is easy to see for the quantum theory, where 
\begin{equation}
 \op{\sigma}_1(x)~=~\int\! \widetilde{\D p}\,
 \left\{ \op{a}(\vec{p})\,\mathrm{e}^{-\I\,p\,x} 
 + \op{b}^\dagger(\vec{p})\,\mathrm{e}^{\I\,p\,x}\right\}\;,
\end{equation}
and the transformation \eqref{eq:SU3CP} corresponds to a map
\begin{equation}
 \mathrm{Out}(\T7)~:\quad
 \op{a}(\vec{p})~\mapsto~\op{a}(-\vec{p})
 \qquad\text{and}\qquad
 \op{b}^\dagger(\vec{p})~\mapsto~\op{b}^\dagger(-\vec{p})\;.
\end{equation}
In contrast, a physical \CP\ transformation of the complex scalar field
$\sigma_1(x)$ would be a map $\sigma_1(x)\mapsto\sigma_1^*(x)$, corresponding to
(see e.g.\ \cite{Branco:1999fs,Sozzi:2008zza}) 
\begin{equation}
 \CP~:\quad\op{a}(\vec{p})~\mapsto~\op{b}(-\vec{p})
 \qquad\text{and}\qquad
 \op{b}^\dagger(\vec{p})~\mapsto~\op{a}^\dagger(-\vec{p})\;.
\end{equation}
It is straightforward to confirm that this na\"{i}ve physical \CP\
transformation of the \T7 states, or any generalization thereof, is not a
symmetry of the action. In fact, this directly follows from the fact that no class--inverting
automorphism exists for \T7. This already shows that physical \CP\
is violated in the \T7 phase.

\medskip

In order see this more explicitly, we construct \CP{}--odd basis
invariants and show that they give rise to an observable \CP\ asymmetry
\begin{equation}\label{eq:asymmetry}
\varepsilon_{\sigma_1\to W\,W^*}~:=~
\frac{\left|\mathscr{M}({\sigma_1\to W\,W^*})\right|^2
-\left|\mathscr{M}({\sigma_1^*\to W\,W^*})\right|^2}{\left|\mathscr{M}({\sigma_1\to W\,W^*})\right|^2
+\left|\mathscr{M}({\sigma_1^*\to W\,W^*})\right|^2}\;,
\end{equation}%
in the decay of a heavy charged scalar $\sigma_1$ to a pair of mutually conjugate heavy gauge boson triplets.
Here $\mathscr{M}({i\to f)}$ denotes the corresponding matrix element.

\medskip

In a perturbative expansion, \CP\ violation arises from interference terms of
diagrams that feature physical \CP{}--odd phases (cf.\ e.g.\
\cite{Branco:1999fs}). Physical observables must be independent of basis choices
for all internal spaces and, therefore, can only depend on basis invariant
quantities. An alternative way to a diagrammatic expansion, thus, is to
construct basis invariants directly. The basis invariant approach is eminently
useful and widely used in the context of \CP\ violation for example in the
standard model \cite{Jarlskog:1985ht, Gronau:1986xb}, but also in extensions
with multiple families \cite{Bernabeu:1986fc}, additional scalars
\cite{Lavoura:1994fv, Botella:1994cs, Branco:2005em, Davidson:2005cw,
Gunion:2005ja, Nishi:2006tg} or for theories with discrete symmetries
\cite{Branco:2015hea, Varzielas:2016zjc}.\footnote{%
See \cite{Fallbacher2015} for comments regarding the conclusions of \cite{Branco:2015hea}.} The reason is that \CP{}--odd
invariants can often be constructed without the need of performing involved
calculations, even if \CP\ violation is arising only at higher loop
order.\footnote{We emphasize that this argument applies to outer automorphisms
in general, not only for the case of \CP{}.} In addition, it has been argued
that the appearance of a single \CP\ odd basis invariant is enough in order to
show that a model is \CP\ violating. To the best of our knowledge, however, it
is not known how basis invariants are related to physical observables in
general. This means that even if \CP\ odd invariants arise in a given model, it
is still a logical possibility that the invariants delicately cancel against one
another in all possible processes. Therefore, in addition to constructing
specific \CP{}--odd basis invariants we also give a specific process for which
we have checked that the invariants do not (all) cancel against one another.

\medskip

\begin{figure}[t]
\centerline{\subfigure[Tree level.]{\CenterObject{\includegraphics{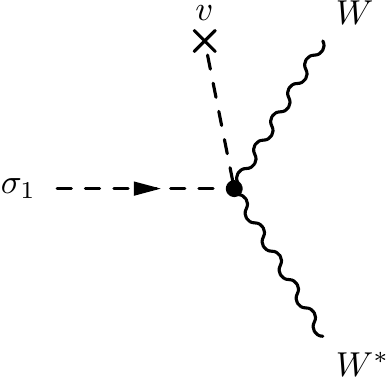}}}
\qquad\subfigure[One--loop.]{\CenterObject{\includegraphics{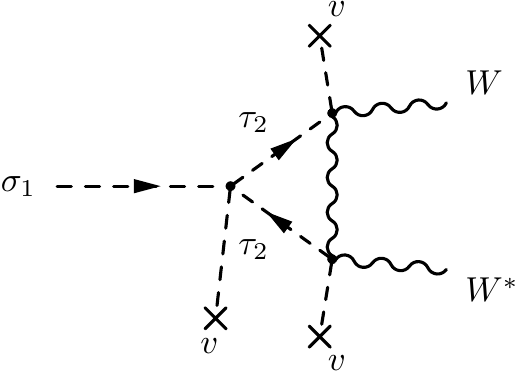}}~\CenterObject{\includegraphics{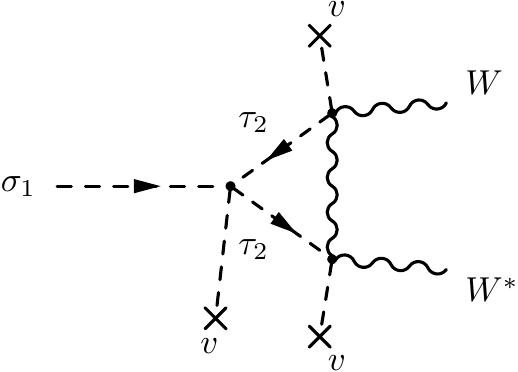}}}}
  \caption{Diagrams whose interference gives rise to \CP\ violation in the decay of $\sigma_1$ to a pair of charged gauge bosons $W$.}
  \label{fig:Decay}
\end{figure}

For the choice of parameters given in equation \eqref{eq:parameters} the
decay $\sigma_1\to W\,W^*$ is kinematically allowed if $g\lesssim0.822$.
The relevant couplings can be obtained by deriving the Lagrangean in the broken phase after the physical fields.
The tree--level coupling of $\sigma_1$ to the charged gauge bosons is given by
\begin{equation}\label{eq:SigmaWW}
 \left[Y_{\sigma_1WW^*}\right]_{ij}~=~\frac{\partial^3\,\mathscr{L}}{\partial\sigma_1\,\partial{W_{\mu,i}}\,\partial{W^*_{\mu,j}}}~\propto~v\,g^2\;.
\end{equation}
Loop corrections to this vertex are possible, for example with triplets
$\tau_2$ in the loop (cf.\ \Figref{fig:Decay}).
The relevant couplings for this correction are given by
\begin{subequations}
\begin{align}\label{eq:SigmaTauTau}
 \left[Y_{\sigma_1\tau_2\tau_2^*}\right]_{ij}~&=~
 \frac{\partial^3\,\mathscr{L}}{\partial\sigma_1\,\partial\tau_{2,i}\,\partial\tau^*_{2,j}}~\propto~v\;,\quad\text{and} \\ \label{eq:TauWW}
 \left[Y_{\tau_2^*WW^*}\right]_{ijk}~&=~
 \frac{\partial^3\,\mathscr{L}}{\partial\tau^*_{2,i}\,\partial{W_{\mu,j}}\,\partial{W^*_{\mu,k}}}~\propto~v\,g^2\;.
\end{align}
\end{subequations}
While $Y_{\sigma_1WW^*}$ can easily be stated in a closed form,
$Y_{\sigma_1\tau_2\tau_2^*}$ and $Y_{\tau_2^*WW^*}$ are in general complicated
functions of the potential parameters $\lambda_i$, see
Equations~\eqref{eq:ySigmaTauTau} and \eqref{eq:yTauWW} in
\Appref{App:Couplings}, where we  also give general expression of the couplings
independently of the potential parameters. From these, in general basis
dependent, couplings it is straightforward to construct basis invariant
quantities via contractions. For that, indices should be contracted such that
basis transformations cancel.\footnote{The basis for each field can, in
general, be rotated independently. However, assuming canonically normalized
kinetic terms, basis transformations cancel in contractions of a field with its
own complex conjugate.} From the given couplings, we find two \CP{}--odd basis
invariant contractions
\begin{align}\nonumber
\mathcal{I}_1~&=~\left[Y^\dagger_{\sigma_1WW^*}\right]_{k\ell}\,\left[Y_{\sigma_1\tau_2\tau_2^*}\right]_{ij}\,
\left[Y_{\tau_2^*WW^*}\right]_{imk}\,\left[\left(Y_{\tau_2^*WW^*}\right)^*\right]_{jm\ell}\;,
\quad\text{and}&
 \\ 
\mathcal{I}_2~&=~\left[Y^\dagger_{\sigma_1WW^*}\right]_{k\ell}\,\left[Y_{\sigma_1
\tau_2\tau_2^*}\right]_{ij}\,\left[Y_{\tau_2^*WW^*}\right]_{i\ell m}\,
\left[\left(Y_{\tau_2^*WW^*}\right)^*\right]_{jkm}\;.
\label{eq:InvariantsI12}
\end{align}
The decay asymmetry $\varepsilon_{\sigma_1\to
WW^*}$ (cf.\ \Eqref{eq:asymmetry}) receives contributions proportional to
$2\,\im\mathcal{I}_{1,2}=\left(\mathcal{I}_{1,2}-\mathcal{I}_{1,2}^*\right)$.
For our choice of parameters one finds
\begin{equation}
 \im\mathcal{I}_1~=~+0.090\,v^4\,g^6\quad\text{and}\quad
 \im\mathcal{I}_2~=~-0.126\,v^4\,g^6\;,
\end{equation}
clearly indicating the presence of \CP\ violation. Inspecting the general
expressions for the invariants \eqref{eq:InvariantsI12} together with the
general expressions of the couplings in \Appref{App:Couplings}, we note that
all contributing complex phases are parameter independent and arise from the
projection of \SU3 Clebsch--Gordan coefficients onto the \T7 subgroup. However,
the geometric phases are weighted by functions of the continuous parameters
$\lambda_i$ of the potential. Therefore, the resulting phases of $\mathcal{I}_1$
and $\mathcal{I}_2$ depend on the potential parameters. Furthermore, we note
that the two invariants are closely related to each other by the relation%
\begin{equation}
 \mathcal{I}_1~=~\omega\,\mathcal{I}_2\;, 
\end{equation}
which is dictated by the $\T7\rtimes\Z2$ symmetry.

\medskip

There are additional contributions to $\varepsilon_{\sigma_1\to WW^*}$ from
other one loop diagrams, for example those containing $\tau_{1,3}$ or gauge
bosons running in the loop. We have explicitly checked that these contributions
do not cancel the asymmetry.

\medskip

Note that all relevant couplings as well as the decay asymmetry are
proportional to positive powers of $v$. Therefore, there is no
physical \CP\ violation when the \SU3 symmetry is restored by taking the limit $v\rightarrow0$.

\section{\boldmath Natural protection of
\texorpdfstring{$\boldsymbol{\theta=0}$}{} in the broken phase\unboldmath}
\label{sec:ThetaProtection}

Finally, let us observe how a possible $\theta$--term \cite{tHooft:1976rip, tHooft:1976snw, Jackiw:1976pf} is affected if \CP\ violation
arises in the way described above. The usual topological term
\begin{equation}\label{eq:Ltheta}
\mathscr{L}_\theta~=~\theta\,\frac{g^2}{32\pi^2}\,G^a_{\mu\nu}\,\widetilde{G}^{\mu\nu,a}\;,
\end{equation}
where $G^a_{\mu\nu}$ is the field strength and
$\widetilde{G}^{\mu\nu,a}:=\frac{1}{2}\varepsilon^{\mu\nu\rho\sigma}G^a_{\rho\sigma}$
its dual, is odd under parity or time--reversal transformations. Therefore,
$\theta=0$ is enforced by the transformation \eqref{eq:CPTrafo} and
$\mathscr{L}_\theta$ is absent from the theory. Crucially, the transformation
\eqref{eq:CPTrafo} is not broken by the VEV. Thus, even though it is not
possible to interpret the transformation \eqref{eq:CPTrafo} as a physical \CP\
transformation of the physical states in the broken phase, the transformation is
unbroken and warrants that $\theta=0$, not only for the broken \SU3, but also for other gauge groups.  We stress that our model is not
realistic and does not even contain fermions. Nevertheless, it is tempting to
speculate that the observed mechanism does provide a symmetry based solution to
the strong \CP\ problem in more realistic settings.

\medskip

For example, one may construct models based on
$[G_\mathrm{SM}\times\SU3_\mathrm{F}]\ltimes\Z2^{\CP}$, where the flavor
symmetry $\SU3_\mathrm{F}$ gets broken spontaneously to \T7, or another type I
group, without breaking $\Z2^{\CP}$. In an intermediate step, one would have
$[G_\mathrm{SM}\times\T7]\ltimes\Z2$, where the \Z2 symmetry continues to forbid
$\theta_\mathrm{QCD}$, while physical \CP\ is violated in the flavor sector. Of
course, the \Z2 must then also be preserved when the remaining \T7 symmetry is
spontaneously broken.  A detailed study of this new avenue in flavor model
building is beyond the scope of this work, but will be explored elsewhere.

\section{Summary and Discussion}
\label{sec:Summary}

We have studied how one obtains a \T7 toy model from a \CP\ conserving  \SU3
theory by spontaneous breaking. \T7 is a so--called type I group, i.e.\ it is
not possible to impose a physical \CP\ transformation on a \T7 model with
generic field content while maintaining the interactions of the theory.
This reflects the fact that there is no basis for \T7 in which all
Clebsch--Gordan coefficients are real.

\medskip

We paid particular attention to the fate of the \SUCP\ transformation. We found
that it is \emph{not} spontaneously broken by the \rep{15}--plet VEV which
breaks \SU3 to \T7. Rather, the \SUCP\ transformation corresponds to the unique
\Z2 outer automorphism of \T7 in the broken phase. This automorphism does
\emph{not} warrant physical \CP\ conservation for the \T7 states, and there is
also no other possible outer automorphism which could do the job. 
Thus, \CP\ is violated in the broken phase.

\medskip

\begin{wrapfigure}[12]{r}[10pt]{7cm}
\CenterObject{\includegraphics{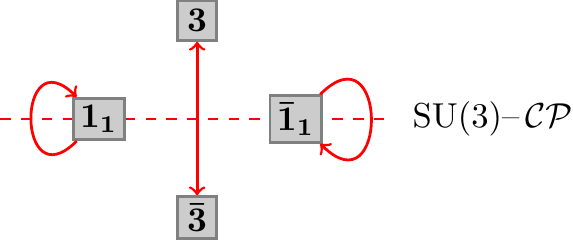}}
\caption{Action of \SUCP\ on the \T7 multiplets contained in the
adjoint of \SU3. A transformation that exchanges both
$\rep{1_1}\leftrightarrow\rep{\bar{1}_1}$ and $\rep{3}\leftrightarrow\rep{\bar{3}}$ is
inconsistent with \T7 and \SU3.}
\label{fig:T7CP}
\end{wrapfigure}

Stated in simple terms, \CP\ is violated because there are \T7 states which
emerge as \emph{complex} linear combinations of certain \SU3 states.  An
example is the $Z$ boson, which transforms as complex \T7 $\rep{1_1}$--plet but
does \emph{not} get conjugated under the \T7 outer automorphism transformation
(cf.\ \Figref{fig:T7CP}).  That is, the outer automorphism cannot be
interpreted as a \CP\ transformation at the level of \T7, and physical \CP\ is
violated. We have demonstrated this explicitly by establishing a decay asymmetry
in the decay of a complex scalar to to massive gauge bosons.

\medskip

Our findings have interesting physical consequences. The definition of
matter and antimatter, at least with respect to a \CP\ mirror, is not
universally possible for chains of groups and subgroups. Rather, the definition
of matter and antimatter depends on the underlying unbroken symmetry. This
has profound implications for cosmology, where the symmetries of the ground
state change in the course of the evolution of the universe.

\medskip

Interestingly, the $\theta$ parameters of \SU3 and other gauge groups remain
forbidden by the outer automorphism, also in the broken phase.  This may allow
one to construct realistic models, in which \CP\ is broken in the flavor sector,
but $\overline{\theta}_\mathrm{QCD}$ is forbidden by the outer automorphism.

\medskip

In our analysis, we have restricted ourselves to only one simple Lie group,
\SU3, and one type I symmetry, \T7. It will be interesting to generalize the
discussion to other groups with richer outer automorphism structure and
include fermions in the discussion.

\acknowledgments

It is a pleasure to thank Mu--Chun Chen for initial collaboration and
discussions, and Renato Fonseca for useful correspondence. We acknowledge
collaboration with Dibya Chakravorty on a related project during the early
stages of this work.  This work has been supported by the German Science
Foundation (DFG) within the SFB-Transregio TR33 "The Dark Universe".

\appendix
\section{\boldmath Details of the quartic \texorpdfstring{$\SU3$}{} invariants\unboldmath}
\label{app:Invarians}
There are in total $14$ linearly independent \SU3 invariants in the contraction $\rep{\bar{15}}\otimes\rep{15}\otimes\rep{\bar{15}}\otimes\rep{15}$.
We use the \SusyNo~\texttt{Mathematica} package
\cite{Fonseca:2011sy} to compute them via the command%
\begin{equation}
 \verb!Invariants[SU3, {{2, 1}, {1, 2}, {2, 1}, {1, 2}}]!\;.
\end{equation}
\SusyNo~provides the invariants ordered according to their permutation group representations and we adopt that ordering.
Only the first five invariants are non--vanishing if all \rep{15}--plets correspond to the same field $\phi$.
In this ordering, the invariants are multiplied by factors $\lambda_i$, $i=1,\dots,5$, resulting in the
quartic part of the potential \eqref{eq:potential}.

\section{\boldmath Details of \texorpdfstring{$\SU3$}{} and the
\texorpdfstring{$\T7$}{}--diagonal basis\unboldmath}
\label{App:SU3}
The basis we choose for the \SU3 triplet generators has been given by Fonseca
\cite{Fonseca:2013qka} and it is implemented in the \SusyNo~\texttt{Mathematica} package
\cite{Fonseca:2011sy}. The generators are given by
\begin{align}\label{eq:TripletGens}\nonumber
 \mathsf{t}^{(\rep3)}_1~&=~\begin{pmatrix} & \frac12 & \\ \frac12 & & \\ & & \end{pmatrix}\;,&
 \mathsf{t}^{(\rep3)}_2~&=~\begin{pmatrix} & & \\ & & \frac12 \\ & \frac12 & \end{pmatrix}\;,&
 \mathsf{t}^{(\rep3)}_3~&=~\begin{pmatrix} &  & \frac\I2 \\& & \\ -\frac\I2 & & \end{pmatrix}\;,& \\
 \mathsf{t}^{(\rep3)}_4~&=~\begin{pmatrix} & -\frac\I2 & \\ \frac\I2 & & \\ & & \end{pmatrix}\;,&
 \mathsf{t}^{(\rep3)}_5~&=~\begin{pmatrix} & & \\ & & -\frac\I2 \\ & \frac\I2 & \end{pmatrix}\;,&
 \mathsf{t}^{(\rep3)}_6~&=~\begin{pmatrix} & & -\frac12 \\ & & \\ -\frac12 & & \end{pmatrix}\;,& \\ \nonumber
 \mathsf{t}^{(\rep3)}_7~&=\frac{1}{2\sqrt{3}}~\begin{pmatrix} 2 & & \\ & -1 & \\ & & -1 \end{pmatrix}\;,&
 \mathsf{t}^{(\rep3)}_8~&=~\begin{pmatrix} & & \\ & \frac12 & \\ & & -\frac12 \end{pmatrix}\;.&
\end{align}
In order to obtain a basis in which the \T7 elements $A$ and $B$ are
block--diagonal we rotate the ($ij$--space) basis of the generators given in
\SusyNo\ according to
\begin{equation}
 \mathsf{t}^{(\rep{15},\T7)}_{a}~=~V^\dagger_{\rep{15}}\,\mathsf{t}^{(\rep{15})}_a\,V_{\rep{15}}\;.
\end{equation}
The corresponding rotation matrix is given by
\setlength\arraycolsep{2pt}
\begin{equation}
 V_{\rep{15}}~=~\tiny
 \frac{1}{\sqrt{3}}\left(
\begin{array}{ccccccccccccccc}
 0 & 0 & 0 & 0 & 0 & 0 & 0 & 0 & 0 & 0 & -\sqrt{2} & 0 & 0 & 1 & 0 \\
 -1 & \mathrm{e}^{-\I\pi/3} & \mathrm{e}^{\I\pi/3} & 0 & 0 & 0 & 0 & 0 & 0 & 0 & 0 & 0 & 0 & 0 & 0 \\
 0 & 0 & 0 & 0 & 0 & 0 & 0 & 0 & 0 & -1 & 0 & 0 & -\sqrt{2} & 0 & 0 \\
 0 & 0 & 0 & 0 & 0 & 0 & \sqrt{3} & 0 & 0 & 0 & 0 & 0 & 0 & 0 & 0 \\
 0 & 0 & 0 & \sqrt{3} & 0 & 0 & 0 & 0 & 0 & 0 & 0 & 0 & 0 & 0 & 0 \\
 0 & 0 & 0 & 0 & 0 & 0 & 0 & 0 & 0 & 0 & 0 & 1 & 0 & 0 & \sqrt{2} \\
 1 &\mathrm{e}^{-2\,\I\pi/3} & \mathrm{e}^{\I\pi/3} & 0 & 0 & 0 & 0 & 0 & 0 & 0 & 0 & 0 & 0 & 0 & 0 \\
 0 & 0 & 0 & 0 & 2\sqrt{2/3} & 0 & 0 & -1/\sqrt{3} & 0 & 0 & 0 & 0 & 0 & 0 & 0 \\
 0 & 0 & 0 & 0 & -1/\sqrt{3} & 0 & 0 & -2\sqrt{2/3} & 0 & 0 & 0 & 0 & 0 & 0 & 0 \\
 0 & 0 & 0 & 0 & 0 & -\sqrt{2} & 0 & 0 & -1 & 0 & 0 & 0 & 0 & 0 & 0 \\
 0 & 0 & 0 & 0 & 0 & -1 & 0 & 0 & \sqrt{2} & 0 & 0 & 0 & 0 & 0 & 0 \\
 0 & 0 & 0 & 0 & 0 & 0 & 0 & 0 & 0 & \sqrt{2} & 0 & 0 & -1 & 0 & 0 \\
 0 & 0 & 0 & 0 & 0 & 0 & 0 & 0 & 0 & 0 & 0 & -\sqrt{2} & 0 & 0 & 1 \\
 0 & 0 & 0 & 0 & 0 & 0 & 0 & 0 & 0 & 0 & -1 & 0 & 0 & -\sqrt{2} & 0 \\
 1 & 1 & 1 & 0 & 0 & 0 & 0 & 0 & 0 & 0 & 0 & 0 & 0 & 0 & 0 \\
\end{array}
\right)\normalsize
 \;.
\end{equation}
\setlength\arraycolsep{6pt}
Due to the degeneracy of \T7 representations in the \rep{15}--plet, we note that
there is a degeneracy in this basis choice corresponding to rotations among \T7
triplets and among anti--triplets, respectively. We have chosen our basis such
that the (anti--)triplet Goldstone modes reside only in one of the
(anti--)triplets, thereby allowing for the simplest form of
\Eqref{eq:unitaryGauge}.

\medskip

In order to obtain a basis for the adjoint space in which the \T7 elements $A$ and $B$ are block--diagonal we rotate the ($a$--space) basis of the generators
according to\footnote{%
The transformation of the gauge bosons is always given by $\left[\mathsf{t}^{(\rep{8})}_{a}\right]_{ij}=f_{aij}$.
}
\begin{equation}
 \mathsf{t}^{(\rep{8},\T7)}_{a}~=~V^\mathrm{T}_{\rep{8},ab}\,\mathsf{t}^{(\rep{8})}_b\;.
\end{equation}
The corresponding transformation matrix is given by
\begin{equation}
 V_{\rep{8}}~=~\frac{1}{\sqrt{2}}\left(
\begin{array}{cccccccc}
 0 & 0 & \I & 0 & 0 & -\I & 0 & 0 \\
 0 & 0 & 0 & \I & 0 & 0 & -\I & 0 \\
 0 & 0 & 0 & 0 & 1 & 0 & 0 & 1 \\
 0 & 0 & 1 & 0 & 0 & 1 & 0 & 0 \\
 0 & 0 & 0 & 1 & 0 & 0 & 1 & 0 \\
 0 & 0 & 0 & 0 & -\I & 0 & 0 & \I \\
 1 & 1 & 0 & 0 & 0 & 0 & 0 & 0 \\
 \I & -\I & 0 & 0 & 0 & 0 & 0 & 0 \\
\end{array}
\right)\;.
\end{equation}
This, of course, is consistent with \eqref{eq:GaugeBosons}, where
\begin{equation}
\left(Z^\mu,{Z^\mu}^*,W^\mu_1,W^\mu_2,W^\mu_3,{W^\mu_1}^*,{W^\mu_2}^*,{W^\mu_3}^*\right)_a^\mathrm{T}~=~\left[V^\dagger_{\rep{8}}\right]_{ab}\,A^\mu_b\;.
\end{equation}

\section{\boldmath Details of \texorpdfstring{$\T7$}{}\unboldmath}
\label{App:T7}

The group \T7 has been used in flavor model building
\cite{Luhn:2007yr,Luhn:2007sy,Hagedorn:2008bc,Cao:2010mp,Cao:2011cp,Luhn:2012bc,Kile:2013gla,Kile:2014kya,Chen:2014wiw}
also motivated by the fact that \T7 is the smallest finite subgroup of \SU3
which has an irreducible triplet representation. \T7 is the smallest
non--Abelian finite subgroup of \SU3 which is not also a subgroup of \SU2 or
\SO3 (cf.\ e.g.\ \cite{Merle:2011vy}). Further details of \T7 can, for example,
be found in \cite{Ishimori:2010au}, whose conventions we follow. \T7 is
implemented in \texttt{GAP} \cite{GAP4} as SmallGroup $\mathrm{SG}(21,1)$.

\medskip

The outer automorphism group of \T7 has order two, and is generated by the transformation
\begin{equation}
 u:~(\elm a,\elm b)\,\mapsto\,(\elm a^6, \elm b)\;.
\end{equation}
The action of the outer automorphism on the irreps has already been given in
\eqref{eq:T7Out}, and it is clearly not class--inverting. We note that the
chosen basis \eqref{eq:T7gens} is an eigenbasis of the outer automorphism,
meaning that the consistency condition
\cite{Holthausen:2012dk,Feruglio:2012cw,Fallbacher:2015rea}
\begin{equation}
 U\,A^*\,U^{-1}~=~A^6\;,\quad U\,B^*\,U^{-1}~=~B\;,
\end{equation}
is solved by $U=\mathbbm{1}$.

\section{Couplings}
\label{App:Couplings}

Here we give general expressions for the couplings defined in equations
\eqref{eq:SigmaWW}, \eqref{eq:SigmaTauTau}, and \eqref{eq:TauWW}. The couplings turn out to be
\begin{subequations}
\begin{align}\label{eq:Ysigma1WW*}
 Y_{\sigma_1WW^*}~&=~\frac{v\,g^2}{\sqrt{6}}\,\mathrm{e}^{-\pi\,\I/6}\,\mathrm{diag}(1,\,\omega,\,\omega^2)\;,
 \\[0.2cm] \label{eq:Ysigma1tautau*}
 Y_{\sigma_1\tau_2\tau_2^*}~&=~v\,y_{\sigma_1\tau_2\tau_2^*}\,\mathrm{diag}(1,\,\omega,\,\omega^2)\;,
\end{align}
\end{subequations}
with the usual $\omega=\mathrm{e}^{2\pi \I/3}$, and
\begin{align}
\left[Y_{\tau_2^*WW^*}\right]_{121}~&=~
\left[Y_{\tau_2^*WW^*}\right]_{232}~=~
\left[Y_{\tau_2^*WW^*}\right]_{313}~=~v\,g^2\,y_{\tau_2^*WW^*}\;,
\end{align}
with 
\begin{equation}
\left[Y_{\tau_2^*WW^*}\right]_{ijk}~=~0\;,
 \end{equation}
for all other choices of indices. 
For a general choice of parameters, the values of the couplings are given by
\begin{align} \nonumber
y_{\sigma_1\tau_2\tau_2^*}~=~&\frac{1}{504\,\sqrt{3}} \left\{ V_{21}^2\, \left[
-14\,\sqrt{10}\, \left(17+5\,\sqrt{3}\,\I\right)\, \lambda_1
+84\, \sqrt{30}\,\left(\sqrt{3}-\I\right)\, \lambda_2 \right.\right. \\ \nonumber
&\left.\hspace{30pt}
{}-240\, \left(1+\sqrt{3}\,\I\right)\,\lambda_4
-\sqrt{10}\,\left(197-55\,\sqrt{3}\,\I\right)\, \lambda_5\right] \\ \nonumber
&{}
+8\,V_{22}^2 \,\left[
28\, \sqrt{10}\,\left(1-\sqrt{3}\,\I \right)\,\lambda_1
-14\,\sqrt{30}\,\I\, \lambda_2+112\,\sqrt{3}\,\I\, \lambda_3\right. \\
\nonumber
&\left.\hspace{30pt}
{}-\left(30-26\,\sqrt{3}\,\I\right)\, \lambda_4
+\sqrt{10}\,\left(20-\sqrt{3}\,\I\right) \, \lambda_5\right] \\ \nonumber
&{} +8\,V_{23}^2\, \left[
28\,\sqrt{10}\, \left(1+\sqrt{3}\,\I\right)\,\lambda_1
-14\,\sqrt{30}\,\I\, \lambda_2
-168\, \lambda_3\right. \\ \nonumber
&\left.\hspace{30pt}
{}+\left(6+65\,\sqrt{3}\,\I\right)\, \lambda_4
-4\, \sqrt{10}\,\left(1-2\,\sqrt{3}\,\I\right)\, \lambda_5\right] \\ \nonumber
&{} +8\,V_{21}\,V_{22}\, \left[
-35\,\sqrt{10}\,\left(1-\sqrt{3}\,\I\right)\,\lambda_1
+21\, \sqrt{30} \left(\sqrt{3}+\I\right)\, \lambda_2 \right. \\
\nonumber
&\hspace{30pt}\left.
{}-56\,\left(3+\sqrt{3}\,\I\right)\,\lambda_3
+6\,\left(1+17\,\sqrt{3}\,\I\right)\, \lambda_4
- \sqrt{10}\left(67+19\,\sqrt{3}\,\I\right)\, \lambda_5
\right]
\nonumber\\
&{} +4\,V_{21}\,V_{23} \left[
-28\,\sqrt{10}\left(2+\sqrt{3}\,\I\right) \,\lambda_1
-42\,\sqrt{30}\left(\sqrt{3}+\I \right)\, \lambda_2 \right. \nonumber\\
\nonumber
&\left.\hspace{30pt}
{}+30\,\left(11+3\,\sqrt{3}\,\I\right)\, \lambda_4
-\sqrt{10}\left(31+11\,\sqrt{3}\,\I\right)\, \lambda_5
\right] \\ \nonumber
&{} -8\,V_{22}\,V_{23}\,
\left[14\, \sqrt{10}\, \lambda_1
-14\,\sqrt{30}\,\I\,\lambda_2\right. \nonumber\\ 
&\left.\left.\hspace{30pt}
{}+10\,\left(3+5\,\sqrt{3}\,\I\right)\, \lambda_4
+\sqrt{10}\,\left(1-3\,\sqrt{3}\,\I\right)\, \lambda_5\right] 
\right\}\label{eq:ySigmaTauTau}
\end{align}
and
\begin{equation}\label{eq:yTauWW}
y_{\tau_2^*WW^*}~=~-\frac{\sqrt{2}}{3}\,\left(2\,V_{21}+V_{22}+2\,V_{23}\right)\;.
\end{equation}
For the choice of parameters given in
\eqref{eq:parameters}, one finds numerical values
\begin{equation}
y_{\sigma_1\tau_2\tau_2^*}~\approx~1.181 + 0.298\,\I
\qquad\text{and}\qquad y_{\tau_2^*WW^{*}}~\approx~-0.295\;.
\end{equation}

\bibliography{Orbifold}
\addcontentsline{toc}{section}{Bibliography}
\bibliographystyle{JHEP}

\end{document}